\documentclass[aps,twocolumn]{revtex4}

\usepackage{amsfonts}
\usepackage{amsmath}
\usepackage{amssymb}
\usepackage{graphicx}
\usepackage{subfigure}
\usepackage{textcomp}

\begin{document}

\title{Extensible router for multi-user quantum key distribution network}
\author{Tao Zhang}
\author{Zheng-Fu Han}
\email{zfhan@ustc.edu.cn}
\author{Xiao-Fan Mo}
\author{Guang-Can Guo}
\affiliation{Key Laboratory of Quantum Information, University of Science and Technology
of China (CAS), Hefei, Anhui 230026, People's Republic of China}
\keywords{QKD network, wavelength routing}
\pacs{03.67.Dd}

\begin{abstract}
A feasible quantum key distribution (QKD) network scheme has been
proposed with the wavelength routing. An apparatus called
\textquotedblleft quantum router\textquotedblright , which is made
up of many wavelength division multiplexers, can route the quantum
signals without destroying their quantum states. Combining with
existing point-to-point QKD technology, we can setup a perfectly QKD
star-network. A simple characteristic and feasibility of this scheme
has also been obtained.
\end{abstract}

\maketitle

Since Bennett and Brassard proposed the famous BB84 protocol in 1984,\cite%
{QKD-BB84} quantum key distribution (QKD) technology has developed greatly. and pear to pear QKD has accessed a practical secure communication system.\cite%
{QKD-B92,QKD-EPR,QKD-experiment-67km-gisin,QKD-experiment-122km,QKD-experiment-stability-ustc,QKD-experiment-125km-ustc}
However, the end to end system can not meet the public need
nowadays, because the internet has embedded in our daily life, so
the QKD network is necessary, and it has become an exigent topic
now.

Then what kind of QKD network should we need? Comparing with the
classical network, we consider that a QKD network should satisfy:
first, it must be a multi-user system and all users are on an equal
footing; second, the coherence of quantum state should be maintained
during the transmission, that means no measurement or amplification
process in the network, especially inside of the router; third, the
security of QKD network should be kept as same as that of pear to
pear QKD system; fourth, the network should be extensible, namely
the users number can be increased or decreased easily without
impacting other users.

A branched network and a looped network proposed by Townsend are the
earliest QKD network schemes,\cite{QKDnet-loopandtree-btlab} in
which the branched network scheme has been demonstrated experimentally.\cite%
{QKDnet-experiment-tree-btlab} These two schemes are simple on
structures and can be easily achieved, but show obviously some
disadvantages. First, when pass through the beam splitters (BS) in
the branched network, photon signals will be delivered to users
randomly, so that some users who do not want signals will get them
and the ones who really want signals will get few. That makes a
great waste of resource. Second, the users are not on an equal
footing in the network, here there is a server who can perform QKD
with any users in the network, but any other two users can not
perform QKD directly. They must do QKD with server first and then
exchange the keys. That may reveal the secure information. Third,
the insertion loss is in proportion to the amount of users. The more
users the network has, the larger insertion loss it will have. This
will restrict the efficient transmission distance of QKD and thereby
reduce the key rate.

Another network scheme proposed by Chip Elliott is a practical one.\cite%
{QKDnet-protocol-bbn,QKD-10nodes-bbn} They use optical switches and
trust relays to build a meshed network. This network is inherently
far more robust than any single point-to-point link since it offers
multiple paths for key distribution. The extensibility of this
network is very good. The efficient QKD distance between two users
can increase by inserting the trust relay station. However, the
trust relay is hazardous for secure information communication, and
the insertion loss will increase in proportion to the amount of
switches, so as to limit the efficient distance too.

A quantum cryptography multi-nodes network system uses wavelength
division multiplex (WDM) technology is disclosed in Japanese Laid
Open Patent Application (JP-P2003-018144).\cite{jppatent} This
system can achieve QKD between any users in the network without
reducing the security of point-to-point system. But it is still a
looped network in which the quantum signals may transit along a
ring, and can not cover a large zone. On the other hand, the
configuration of each node is different and complicated, and it has
not been fulfilled till now. To my knowledge, all of these schemes
can not match total requirements above.

Here we present a new QKD network scheme which also utilizes WDM
technology. But the center of this network is a \textquotedblleft
quantum router\textquotedblright, which deliver the different
quantum signal to the appointed node of the system according to its
wavelength, and the quantum states of signals will not be destroyed
during this process. Its function is same as a traditional router
but the mechanism is different. This router can be comprised of some
wavelength division multiplexers (MUXs) easily.

\begin{figure}[tbp]
\centering
\includegraphics[width=\columnwidth]{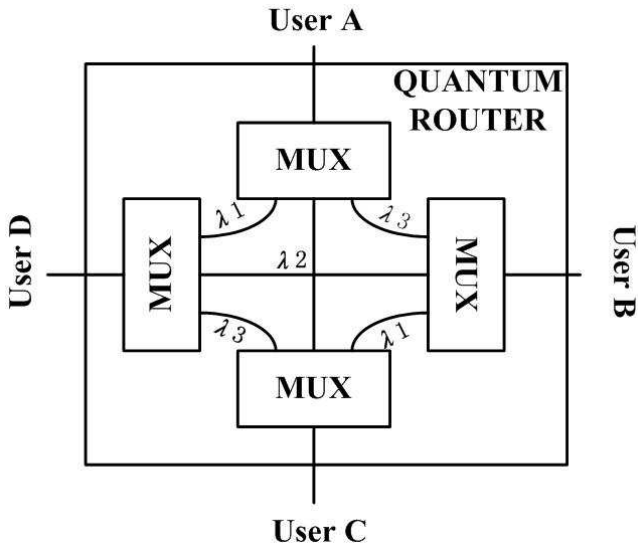}
\caption{topography of a four ports quantum router} \label{fig1}
\end{figure}

FIG. \ref{fig1}
is a block diagram of a 4 ports \textquotedblleft quantum
router\textquotedblright\ which is made up of 4 MUXs. Each MUX has 3
demultiplexing ports and 1 multiplexing port. All demultiplexing
ports of 4 MUXs connect with each other by fibers. The connection
scheme satisfy that any two MUXs have one and only one link which
transfer same wavelength signals. Each multiplexing port connects
with one user by arterial fiber. All users and the \textquotedblleft
quantum router\textquotedblright\ compose a QKD star-network. This
type of network with any amount of users can also be setup. The
connection scheme of MUXs accords with edge coloring theorem in
graph theory.\cite{Graph-theory} Each MUX corresponds to a vertex of
graph, each link corresponds to an edge, and each wavelength
corresponds to a color. The edge coloring theorem says: for a
complete graph with N nodes, when N is even, it needs N-1 colors to
color each edge so that adjacent edges (edges which has same vertex)
have different colors; when N is odd, the amount of colors is N.
That means, for a N ports \textquotedblleft quantum
router\textquotedblright , each MUX needs N-1 or N demultiplexing
ports to finish the connection.

\begin{figure}[tbp]
\centering \subfigure[N=5]{\label{fig2a}\includegraphics{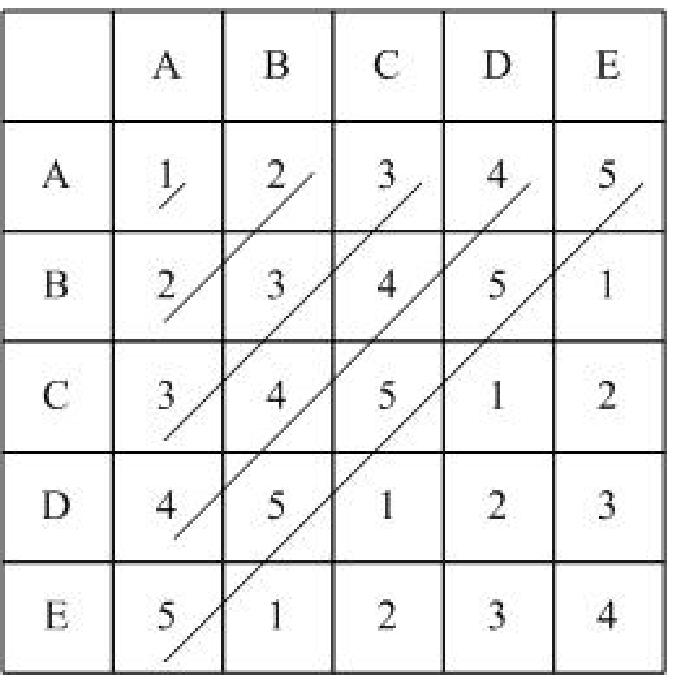}}
\subfigure[N=6]{\label{fig2b}\includegraphics{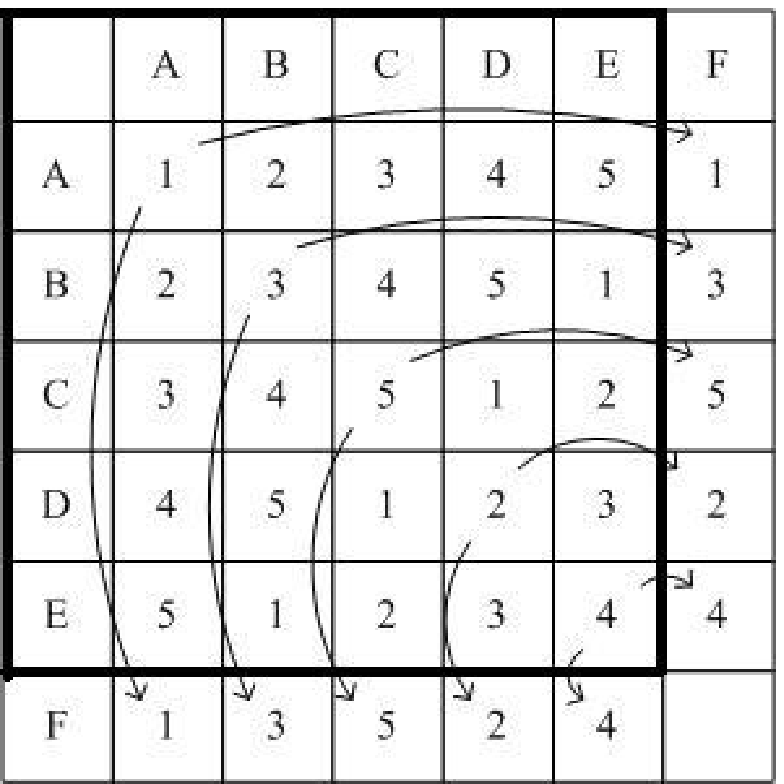}}
\caption[connecting relation between N MUXs]{connecting relation
between N MUXs (A,B,C\ldots represent MUXs, 1,2,3\ldots represent
wavelengths, the number 2 in row A and in line B means connect A and
B by wavelength 2)}
 \label{fig2}
\end{figure}

Here we give a simple way to find the right connection scheme for
odd or even nodes router. As shown in FIG. \ref{fig2},
the capital letters represent MUXs and Arabic numerals represent
wavelengths. The tables in the figure indicate the connecting
relation between MUXs. For example, the number 2 in row A and in
line B means connect A and B by wavelength 2. The process of
finishing these tables is: when N is odd, e.g. N=5 such as in
fig.\ref{fig2a}, fill the blanks along 45\textdegree\ diagonal with
numbers 1 to N in turn, and when it reaches N, go back to 1 and fill
the rest blanks just as above; when N is even, e.g. N=6 such as in
fig.\ref{fig2b}, use said method to fill a N-1 table first, and then
copy 135\textdegree\ diagonal numbers to last row from left to right
and last line from up to down.

Because this \textquotedblleft quantum router\textquotedblright\ is
made up of MUXs, its performance depends on that of MUXs. Up to
date, the arrayed waveguide gratings (AWG) and thin film
interference filters WDM are gaining prominence. Filters offer good
stability and isolation between channels at moderate cost, but with
a high insertion loss. AWGs exhibit a flat spectral response and low
insertion loss, and they are also better for large channel counts,
where the use of cascaded thin film filters is impractical.

For one thing, MUXs' capability will impact the size of network. The
maximum nodes of a network depends on the max amount of channels of
MUX. The popular dense WDM product has 40 channels, and 4200
channels have been achieved in laboratory.\cite{WDM-4200-channels}
That means it is possible to build a \textquotedblleft quantum
router\textquotedblright\ with 4200 ports in future. Consider that
the absolutely secure communication is not as popular as internet,
this QKD net size may be enough for a big city.

Secondly, the insertion loss will reduce the efficient QKD distance.
Since signals will pass through MUX twice when they pass the
\textquotedblleft quantum router\textquotedblright , the insertion
loss of \textquotedblleft quantum router\textquotedblright\ is
double. The insertion loss of popular product is 5 dB, so it is 10
dB of \textquotedblleft quantum router\textquotedblright . According
to the performance of present point-to-point QKD system, we can
build a QKD network over 50 km at least. That will still meet the
requirement of a big city. Along with the development of WDM
technology, the insertion loss will be less than 1 dB in
future,\cite{WDM-AWG-progress} and then the quantum network will
cover more than 100 km.

The third problem is crosstalk. For a network, crosstalk will bring
bit errors, so it must be reduced as low as possible. We think it
can be estimated as follow: The insertion loss ($IL$) and crosstalk ($FC$) is:%
\begin{equation}
IL=10\times \log (P_{in}/P_{out})  \label{equ1}
\end{equation}%
\begin{equation}
FC_{j}(\lambda _{i})=10\times \log [P_{j}(\lambda _{i})/P_{i}(\lambda _{i})]
\label{equ2}
\end{equation}%
Here $P_{in}$ and $P_{out}$ are input and output probability of
single photons, $P_{j}(\lambda _{i})$ is output probability of
photons with wavelength $\lambda _{i}$ which export from
port j, $P_{i}(\lambda _{i})$ is output probability of photons with wavelength $%
\lambda _{i}$which export from port i. $P_{i}(\lambda _{i})$ in equation \ref%
{equ2} equal to $P_{out}$ in equation \ref{equ1}
\begin{equation}
\frac{P_{out}}{P_{in}}=10^{-IL/10}  \label{equ3}
\end{equation}%
\begin{equation}
\frac{P_{j}(\lambda _{i})}{P_{in}}=\frac{P_{j}(\lambda
_{i})}{P_{out}}\times \frac{P_{out}}{P_{in}}=10^{[FC_{j}(\lambda
_{i})-IL]/10}  \label{equ4}
\end{equation}%
We first assume that all input probability of single photons from
any user is same when they enter the \textquotedblleft quantum
router\textquotedblright, since one photon will pass through two
MUXs when it pass \textquotedblleft quantum
router\textquotedblright, crosstalk versus efficient signals is:
\begin{equation}
\frac{[P_{j}(\lambda
_{i})/P_{in}]^{2}}{(P_{out}/P_{in})^{2}}=10^{2\times FC_{j}(\lambda
_{i})/10}  \label{equ5}
\end{equation}%
Now we consider the situation that the probability of input photons
is not the same. A terrible situation is that input photons which
produce efficient signals pass through a fiber that has X dB
insertion loss before pass through \textquotedblleft quantum
router\textquotedblright\ but those photons which produce crosstalk
do not. The ratio in equation \ref{equ5} will become $10^{[X+2\times
FC_{j}(\lambda _{i})]/10}$. If there are
many inputs that produce crosstalk, the ratio must be $\sum_{j=1,j%
\neq i}^{N-1}10^{[X+2\times FC_{j}(\lambda _{i})]/10}$, here $N$ is
the amount of channels. For popular product, $N=40$, $FC_{j}(\lambda
_{i})<-25dB$ (when $j=i\pm 1$), $FC_{j}(\lambda _{i})<-30dB$ (when
$j\neq i\pm 1$), $X<10$. The ratio is less than 0.056\%. So the
errors brought by crosstalk are less than other affect and can be
ignored. Along with the development of WDM technology, crosstalk
will be smaller and the performance of \textquotedblleft quantum
router\textquotedblright\ will improve.

In this paper we provide a new scheme of QKD network. The kernel part of
this network is \textquotedblleft quantum router\textquotedblright\ which is
made up of MUXs. Base on present WDM technology and combine with
point-to-point QKD system, we can easily build a feasible QKD network over
50 km with 40 users.

This work was funded by National Knowledge Innovation Program piloted at
CAS, National Natural Science Foundation of China (60537020) and National
Science Foundation for Innovative Research Group (60121503).

\bibliographystyle{apsrev}
\bibliography{myqkdref}

\end{document}